%% file: main.tex
\documentclass[10pt,journal,letterpaper]{IEEEtran}

\usepackage{cite}

\ifCLASSINFOpdf
\else
\fi

\usepackage[cmex10]{amsmath}
\usepackage{amsfonts}

\usepackage{algorithm}
\usepackage{algorithmic}
\usepackage[normalem]{ulem}

\usepackage{graphicx}
\usepackage[T1]{fontenc}
\usepackage{geometry}

\begin{document}
%

\title{Pushing towards the Limit of Sampling Rate: Adaptive Chasing Sampling}




\author{ {
$^\dagger$Ying Li, $^\dagger$$^\ddagger$Kun Xie, $^\dagger$Xin Wang}
\\ $^\dagger$Dept of Electrical and Computer Engineering, Stony Brook University, USA
\\ $^\ddagger$College of Computer Science and Electronics Engineering, Hunan University, Changsha, China
\\Email: yingli@ece.stonybrook.edu, xiekun@hnu.edu.cn, xwang@ece.stonybrook.edu
\thanks{\copyright 2015 IEEE. This paper has been accepted to MASS 2015. Personal use of this material is permitted. Permission from IEEE must be obtained for all other uses, including reprinting/republishing this material for
advertising or promotional purposes, collecting new collected works for resale or
redistribution to servers or lists, or reuse of any copyrighted component of this work in other
works.}
}
\maketitle

\input abstract

\input Sec1_Introduction

\input Sec2_Related

\input Sec3_Background

\input Sec4_SystemOverview

\input Sec5_MainScheme

\input Sec6_Simulation

\input Sec7_Conclusion

\bibliographystyle{abbrv}
\bibliography{myref}
%

\end{document}

%% file: abstract.tex
\begin{abstract}
Measurement samples are often taken in various monitoring applications. To reduce the sensing cost, it is desirable to achieve better sensing quality while using fewer samples. Compressive Sensing (CS) technique finds its role when the signal to be sampled meets certain sparsity requirements. In this paper we investigate the possibility and basic techniques that could further reduce the number of samples involved in conventional CS theory by exploiting learning-based non-uniform adaptive sampling.
 Based on a typical signal sensing application, we illustrate and evaluate the performance of two of our algorithms, Individual Chasing and Centroid Chasing, for signals of different distribution features. Our proposed learning-based adaptive sampling schemes complement existing efforts in CS fields and do not depend on any specific signal reconstruction technique. Compared to conventional sparse sampling methods, the simulation results demonstrate that our algorithms allow $46\%$ less number of samples for accurate signal reconstruction and achieve up to $57\%$ smaller signal reconstruction error under the same noise condition.
\end{abstract} 

%% file: Sec1_Introduction.tex
\section{Introduction}
\label{Section:Introduction}
Efficient information collection is critical for many applications, such as medical imaging, radar detection and spectrum sensing. Practical signals are generally continuous and can be sampled into digital form for more efficient storage, processing and communications. Obviously, a higher number of samples would lead to larger resource consumption and higher processing complexity. The fundamental challenge is to achieve the desired degree of signal fidelity in an often noisy environment with the minimum number of samples.

Compressive Sensing (CS)~\cite{CSCandesTao1, CSCandesTaoRomberg1, CandesRomberg1, CSDonoho} has attracted a lot of recent attention, with its capability of reconstructing sparse signals with the number of samples much lower than that of the Nyquist rate.
The fundamental works of CS include the introduction of the $l_1$-minimization to reconstruct the signal, and more recently the greedy recovering of signal components gradually~\cite{OMP1}\cite{SwOMP}\cite{ROMP}\cite{TargetCounting}. Attempts have been made to directly apply CS in different application areas, including the reduction of traffic volume during signal acquisition~\cite{SensorCS}\cite{RandomizedGossip}, and finding target locations and numbers~\cite{TargetLocalization}\cite{TargetCounting}\cite{RSSI_CS}.

Given the limitation of sensor resources and battery energy, it is often desirable to minimize the number of sensors involved in a single sensing task. This will in turn contribute to a lower average duty cycle of all the sensors and the extension of the reliable working life of the sensor network. In applications that require the participation of surrounding devices, for example the fast growing research field of crowd sensing, certain rewards can be given to the contributors to motivate their involvement and cooperation in sensing. In these cases, minimizing the number of involved sensors helps reduce the extra sensing cost besides power consumption.
%

Conventional compressive sensing schemes take samples randomly and uniformly in a sampling space. This inevitably makes it difficult to determine the minimum number of samples to take. If the samples are not sufficient, the signal cannot be accurately recovered. Thus to be safe, usually more than enough samples are taken to guarantee the accurate reconstruction at higher cost. Instead of taking samples all at once, we would like to find out if the samples can be taken sequentially, with subsequent number of samples determined based on the estimation results from previous samples. More specifically, we would like to find ways of adapting the sensing matrix that defines the sensing behaviors based on the information learnt from previous observations to guide subsequent measurements to "focus" on the areas detected with possible existence of signals, so that the overall number of samples can be greatly reduced. Although it may incur some delay in collecting samples over multiple rounds, this method may help significantly reduce the sensing resources when there is a need to continuously monitor some targets or signals, while the chance of appearance of the targets or signals is very small.

  Fundamentally different from some existing CS algorithms~\cite{CSCandesTao1, BasisPursuit, l1Norm, OMP1} whose aims are to improve the quality of reconstructed signals with a given number of samples, our proposed algorithms aims to reduce the number of samples needed for the whole sensing process through the adaptation of sampling distribution, taking advantage of the learning process to achieve similar or even better signal recovery accuracy. Different from conventional schemes that search for signals in random locations which may miss the signals, sensors selected in our scheme first detect the potential existence of signals although their received values may be low, and the knowledge is then applied to guide efficient signal sensing later.


To illustrate the principle of our proposed learning-based adaptive CS schemes, we construct a sensing matrix according to a practical application, which allows us to map the change of the sensing matrix in theory to the choice of sensor measurement locations in practice. Specifically, we consider the detection and estimation of the strengths and locations of sparsely distributed signal sources using sensors embedded in commonly accessible wireless devices such as cell phones, which have been considered for use in crowd sensing. Detection of the signal
strength on different locations finds its use in many applications, including the emerging cognitive radio field where cell phones can be used collaboratively to form the spectrum sensing maps on the activities of primary users, localizing the sound sources for detecting events such as gunshot or riots, and detection of radiation sources. Our adaptive sampling scheme can be applied to broad categories of energy signal sensing.


We propose two learning-based adaptive sensing algorithms for different application scenarios to significantly improve the sensing efficiency and quality. Our algorithms do not depend on the underlying CS reconstruction methods, thus can be applied in many CS-suitable applications. The features of our work can be summarized as follows:
\begin{itemize}
  \item Compared to conventional schemes which employ completely random sampling, our adaptive algorithms wisely choose where to sample based on the knowledge from previous sampling process, thus requiring much fewer samples to estimate the signals at desired accuracy level.
  \item Compared to a few existing adaptive sensing schemes~~\cite{DataGathering, CSEnergy}  in the literature:
  \begin{itemize}
    \item The literature studies take samples uniformly although the sampling rate is adapted in the temporal or spatial domain to improve the sensing quality, while our schemes adapt the sample distribution based on the learning from previous estimation results to greatly reduce the total number of samples needed.
    \item Our unique sensing matrix structure allows convenient reuse of samples previously taken along with later samples for higher sensing accuracy while avoiding tasking same sensors multiple times to conserve sensing resources and reduce the overhead in transmitting sensing signals in the network.
  \end{itemize}
\end{itemize}
As we will  theoretically show later, if a sensor in our schemes is tasked once, it only takes measurement in one round in the whole adaptive sensing process, thus ``the number of sensors used'' equals ``the number of samples'' in our work, and we refer to them inter-changeably in the paper. Therefore, minimizing the number of samples is equivalent to reducing the number of involved sensors, which helps to conserve the sensing cost involved for both sampling and communications. In addition, our schemes achieve much lower reconstruction error compared to conventional CS schemes under the same level of noise for the same number of sensors used.

The rest of the paper is organized as follows. In Section~\ref{Section:RelatedWork}, some related works are discussed. Section~\ref{Section:Background} gives the fundamentals of CS. Our system model and formulation are described in Section~\ref{Section:SystemOverview}. Section~\ref{Section:MainScheme} introduces the two algorithms of adaptive sensing. We evaluate the performance in Section~\ref{Section:Simulation}, and conclude the work in Section~\ref{Section:Conclusion}.

%% file: Sec2_Related.tex
\section{Related Work}
\label{Section:RelatedWork}
The research problems on CS can be roughly divided into two categories. One is to study and improve the reconstruction technique itself in order to reduce the computational complexity or increase the recovery accuracy against noises. Another is the application of CS to solving practical problems.

For signal reconstruction, the $l_1$-minimization algorithm \cite{CSCandesTao1, BasisPursuit, l1Norm}, also called basis pursuit, transforms the original reconstruction problem into a linear programming problem to solve with convex optimization; Another family of methods is greedy-based, including Orthogonal Matching Pursuit (OMP) \cite{OMP1}, Stagewise OMP (StOMP)\cite{SwOMP} and Regularized OMP (ROMP) \cite{ROMP}, which tackles the reconstruction problem by gradually recovering the components of the signal in each iteration. The $l_1$-minimization is generally believed to give better reconstruction performance. While the greedy-based methods have the advantages of simpler implementation and lighter computational overhead, they require more measurements for the same reconstruction quality. Other reconstruction algorithms include iterative thresholding methods \cite{IterativeThresholding1, IterativeThresholding2} and various Bayesian methods\cite{BayesianCS1, BayesianCS2}.

CS has been directly applied in various network and sensing applications. In the network area, some schemes have been proposed to reduce the total number of network messages by aggregating and forwarding the linearly weighted summation of the messages on the path towards the destination \cite{SensorCS} or a neighbor~\cite{RandomizedGossip}, taking advantage of the capability of CS to recover the original messages.
The Greedy Matching Pursuit (GMP) algorithm for target counting in \cite{TargetCounting} assumes that the nonzero values can only take several possible integers and exhaustively try out each position of the vector. This method is too computational costly for signal vectors of large dimensions, and nearly impossible for vectors with arbitrary values.
\cite{RSSI_CS} straightforwardly adopts CS into the Access Point received signal strength based localization problem. CS also finds its use in the field of Cognitive Radio recently. The work in \cite{CR} exploits CS to estimate the occupied spectrum channels and the locations of Primary Radios.


Iterative CS algorithms have also been proposed in the literature. In the field of image/video processing where the computing overhead is not a constraint, multiple rounds of CS reconstructions are sequenced to achieve better quality of recovered signal accuracy~\cite{MRI}\cite{UnderSamplingMRI}
at the cost of higher number of samples and computation overhead. There are also attempts to adaptively reconstruct components of the signals for an overall better performance~\cite{adaptive1, adaptive2, adaptive3}. Despite the increase of  tolerance to noise, these adaptive CS algorithms cannot be applied to effectively reduce the number of samples.



In the existing CS theory, the necessary number of samples to take is given and derived for basic $l_1$ minimization based on random sampling.  In this work, we investigate the possibility and ways of reducing the number of samples needed as well as the computational complexity while maintaining the sensing accuracy. Rather than taking random samples as in conventional CS schemes, to  reduce the number of samples and improve the recovery quality, we propose to adapt the sensing matrix taking advantage of the iterative learning process.

%% file: Sec3_Background.tex
\section{Fundamentals of Compressive Sensing}
\label{Section:Background}
Conventional information theory mandates a sampling rate to be at least twice the bandwidth of the signal being sampled. Recent research shows that a sparse signal can be reconstructed through Compressive Sensing with a high probability at much lower sampling rate. Moreover, most signals that are not sparse enough can also be projected to other domains to achieve the desired sparsity.

Let vector $x \in \mathbb{R} ^N$ be a signal not sparse enough. Given an $N \times N$ orthogonal basis $\Psi=[\Psi_1, \Psi_2,..., \Psi_N]$ with each $\Psi_i$ being a column vector, we have:
\begin{equation} x=\Psi s = \sum_{i=1}^{N} s_i\Psi_i \end{equation}
where $s$ is the coefficients of $x$ in the transformed domain $\Psi$.
$s$ is said to be $k$-sparse if it has at most $k$ nonzero entries and $k\ll N$. The samples are then
\begin{equation}y=\Phi x = \Phi\Psi s = As \label{eq:xAs}\end{equation}
where $\Phi$ is an $M \times N$ measurement matrix to be defined later with $k \ll M \ll N$, the combined $M \times N$ matrix $A$ is called the sensing matrix, and $y$ is the sample vector of $M \times 1$.

Under the condition of $l_0$ minimization: $\min\|\mathbf{s}\|_{l_0}$, instead of acquiring $N$ samples of $s$,  only $M=2k$ of noise-free measurements are needed to reconstruct $s$~\cite{M2k}. However, this problem is NP-hard. If $A$ meets the Restricted Isometry Property (RIP) condition,  it is much easier to solve the $l_1$-minimization problem below

\begin{align}
\min\|\mathbf{s}\|_{l_1} \\
{\mbox{s.t. }} \hspace{0.05in} \|\mathbf{A}\mathbf{x}-\mathbf{y}\|_{l_2} \leq\varepsilon
\end{align}

where the parameter $\varepsilon$ is the bound of the error, as long as $M \geq c \cdot k \cdot log (N/k)$~\cite{CSCandesTaoRomberg1}.

Nonetheless, the lower bound requirement of $M=2k$ measurements is rarely achieved using conventional CS methods. The major contribution of this paper is to provide some insights on the possibility and potential strategies that can be exploited to push the sampling rate towards the limit.

%% file: Sec4_SystemOverview.tex
\section{Problem Formulation}
\label{Section:SystemOverview}
To investigate the possibility and methods of improving the sampling efficiency and accuracy, in this paper, we instantiate a specific problem of using cognitive cell phones which can switch to other spectrum bands to detect the strength and locations of primary signal sources for the potential cognitive use of the unoccupied spectrum of the measured bands. This setting helps to picture the mapping of theoretical changes on the sensing matrix into the adjustment of sensor positions/sampling points in practice.

We consider a general scenario where some cell phone users are willing to participate in estimating the strength and locations of aggregate primary radio signals (e.g., TV or radar signals) which are generally located sparsely in the sensing field.
Since we utilize cell phones to detect the signal strength, we consider cell phones as spectrum "\textbf{sensors}" in this paper.
To provide a location reference and facilitate the scalable monitoring, the sensing domain is partitioned into $N$ grids with the size determined based on the resolution requirement of a specific application, and each grid could have no or several signal sources inside it. Similarly, a grid could have no or several cell phones, each can update the base station its grid location upon grid crossing by piggying back the information with other uplink messages.

While some literature work \cite{TargetCounting} assume that signal sources have identical transmitting power to count the number of targets, we do not have this restriction or to differentiate individual signals within a grid. Instead, we consider signals within a grid as an \textbf{aggregate signal source} located at the grid center.
Generally, within the short sensing period, we don't expect a significant change on the strength of the signals to be measured and a large number of signal sources or sensors to move across grids.

\begin{figure}[h]
\centering
\includegraphics[width=2.8in]{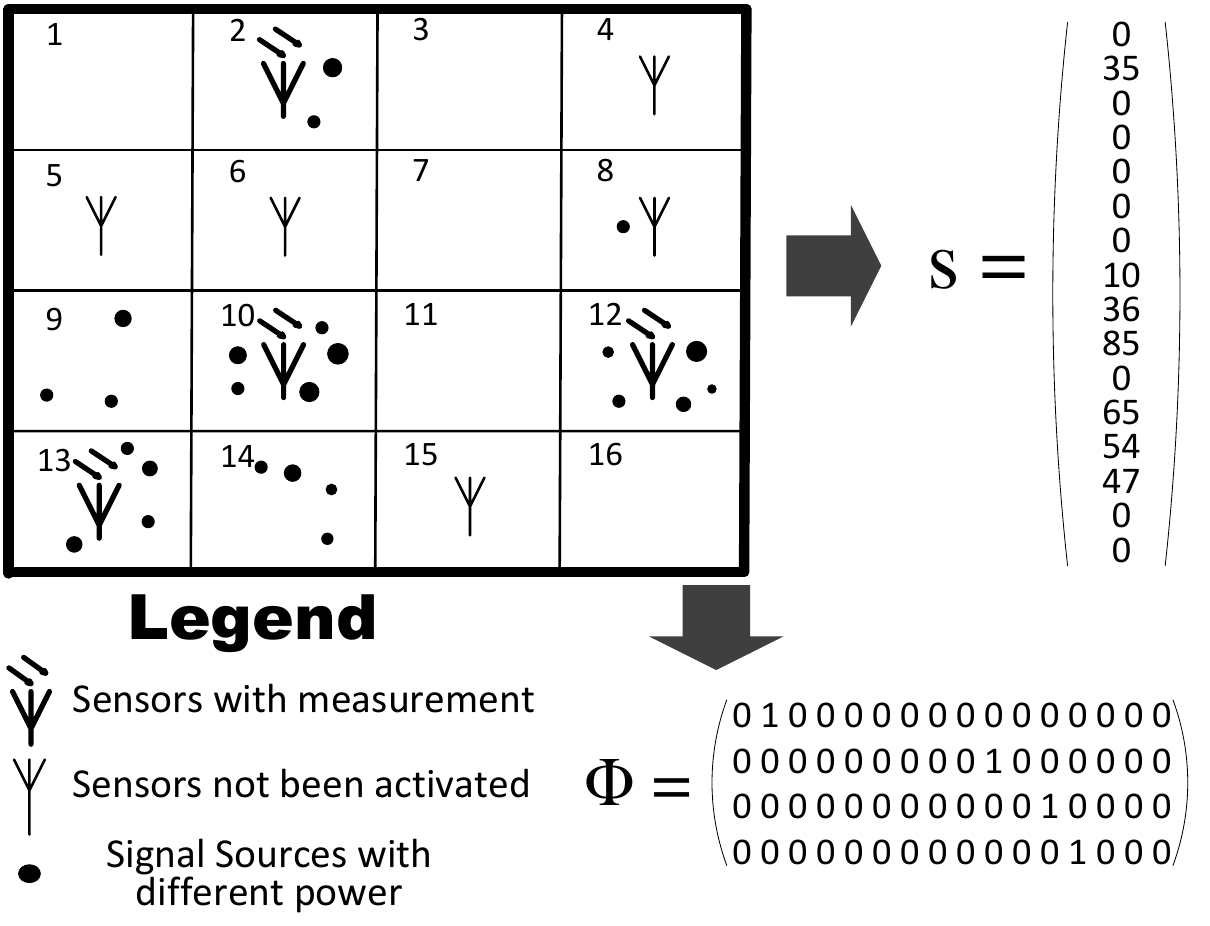}
\caption{A demo of the system with the corresponding signal vector $s$ and measurement matrix $\Phi$.}
\label{fig:Formulation}
\end{figure}

Figure~\ref{fig:Formulation} shows an example system of 16 grids in a monitoring region. Signal sources with different level of energy are indicated by the dot of different sizes.
For grids that have several phones inside, we can always pick only one cell phone to do the task in order to preserve the sensing resources. Thus it is safe to assume each grid has at most one sensor for a specific sensing task at a given time. Cell phones tasked for sensing the spectrum will send their samples to the base station to be fused.

To monitor the energy of signals, traditionally, a large number of sensors are placed 
across the whole monitoring domain and kept active to maintain the coverage~\cite{TraditionalLocalization1}\cite{TraditionalLocalization2}. In reality, for a given sensing resolution, i.e. grid size in this setup, only a small number of grids will contain signal sources at a given time. This spatial domain sparsity makes it possible to apply the compressive sensing technique to reconstruct the locations and strength of the signal sources.

Instead of tasking an adequate number of sensors~\cite{TargetCounting} randomly based on the CS recovery requirement, to increase the recovery accuracy while reducing the total number of sensors involved, we would like to start with a small number of sensors at random locations. We then adaptively task additional sensors by learning from past measurement data. Next we will show mathematically how our matrix structure helps to facilitate this process.

The strength or energy of many real-world signals decays over the distance, and could also be impacted by the environment. Our proposed schemes do not rely on any specific channel models, therefore we denote the channel factor between the radio transmitter in grid $i$ and receiver in grid $j$ as $C_{ij}$. The channel fading and environment impact often add in random factors into $C_{ij}$, rendering the channel to be Gaussian.


We abstract the distribution of signal sources into vector presentation. Let $s = [s_{1}, s_{2}, ..., s_{N}]^T$ be an $N \times 1$ column vector, where the $i^{th}$ entry $s_{i}$ is the aggregate signal strength of grid $i$. $s$ is $k$-sparse with $k\ll N$, which means at most $k$ grids out of $N$ actually have signal sources.



Let $\Psi$ be an $N \times N$ transformation matrix which embodies the signal energy decaying process over the radio channel:
\begin{equation} \Psi =
\begin{pmatrix}
  C_{11} & C_{21} & \cdots & C_{N1} \\
  C_{12} & C_{22} & \cdots & C_{N2} \\
  \vdots & \vdots & \ddots & \vdots \\
  C_{1N} & C_{2N} & \cdots & C_{NN}
\end{pmatrix} \label{eq:Psi}\end{equation}
Then $x=\Psi s$ is the received signal strength vector with $x(j)$ denoting the aggregate signal strength received by the sensor at grid $j$ from all signal sources.

As illustrated in Figure~\ref{fig:Formulation}, the sensors tasked to take samples can be "selected" by an $M \times N$ measurement matrix $\Phi$. The $m^{th}$ row of the matrix is a $1 \times N$ row vector with all elements equal to zero except $\Phi(m,j) = 1$, where $j$ is the index of the grid at which the sensor $m$ is located. Each entry $x(j)$ denotes the signal strength received at the sensor in grid $j$, so the effect of left-multiplying $\Phi$ with the vector $x$ in Equation(\ref{eq:xAs}) is to select $M$ out of $N$ rows of $x$, or equivalently choose the $M$ sensors at specific grid positions to take samples.

Under this formulation, a set of sensors can be flexibly and conveniently specified with matrix $\Phi$ to take samples in each round of the adaptive sensing process.
We can take advantage of the samples acquired in the previous iterations of the adaptive process and combine them with the new samples to form a more informative sample vector $y$, where
\begin{equation}
y=
\begin{pmatrix}
y' \\
y''
\end{pmatrix}
=\Phi \Psi s
=
\begin{pmatrix}
\Phi' \\
\Phi''
\end{pmatrix}
\Psi s
\label{eq:MatrixReConstruction}
\end{equation}
to recover the data using the new $\Phi$ matrix. In Equation~\ref{eq:MatrixReConstruction}, $y'$ is the vector of samples already collected by sensors previously selected by $\Phi'$, and $y''$ contains the new samples taken by sensors newly specified in $\Phi''$. A selected sensor only needs to take sample once during the whole sensing process, with its sample saved and used with later samples to reconstruct the signal. Therefore, the total number of samples taken in the whole sensing period equals the number of sensors used, thus we do not differentiate the number of samples from the number of sensors thereafter. The spectrum map can be updated periodically, and for each period, a new sequence of sensing can be initiated.
The reuse of samples from previous rounds is enabled by our unique sensing matrix structure, which helps to greatly reduce the sensing resource consumption and serves as the basis of our algorithms to be proposed.

The sensing matrix $A = \Phi \Psi$ has been proven in \cite{CSDonoho,TargetCounting} to obey RIP condition as long as matrix $\Phi$ and $\Psi$ are constructed as defined above. Therefore in each adaptive iteration, we can safely apply the $l_1$-minimization method to perform the reconstruction. Although the small number of sensors used in the beginning phase of the adaptive process would result in inaccurate reconstruction, in the next section, we will show that by learning the result from each previous iteration, the process eventually converges to an accurate reconstruction, while the overall number of sensors used is significantly reduced.

%% file: Sec5_MainScheme.tex
\section{Adaptation of Sensing Matrix and Process for Improved Sensing Quality}
\label{Section:MainScheme}




In this work, we propose two learning-based methods in which a small set of sensors considered to be better will be tasked to take samples in each round of sensing based on the reconstruction results from the previous round, then the new samples together with the existing samples will generate an improved reconstruction, and so forth. In this section, we first present our basic setting in Section~\ref{Setting}, and then introduce in details of our sensing algorithms in Section~\ref{Individual} and ~\ref{Centroid}, respectively. Finally, we discuss our strategy in avoiding the local optimal solution in Section~\ref{Local_Optimum}.

\subsection{Basic Setting and Design Consideration}
\label{Setting}
In this paper, we use terms such as "task", "turn on" or "add" when selecting a sensor to take measurement. 
The locations of sensors in terms of which grid are denoted by the set $\mathbb{L}$.
For example, in Figure~\ref{fig:Formulation} sensors are installed at grids 2, 4, 5, 6, etc., then $\mathbb{L}$ will have the corresponding elements.

As an iterative scheme, the algorithm needs an initial value to start with. Specifically, at the beginning of our adaptive sensing process, we need to decide the number of sensors to take samples for the first round, then the adaptive algorithm will handle the following iterations automatically. In order to determine the optimal number of samples to take, conventional CS techniques often assume the prior knowledge of the sparsity value $k$ or estimate $k$, and their performances are highly impacted by the accuracy of $k$ value taken. In contrast, our schemes take samples sequentially and determine the number of additional samples needed based on the previous estimation results, and the initial $k$ is only used as a reference.
As mentioned in Section~\ref{Section:Background}, $2k$ ($k$ is the sparsity) is believed to be the theoretical extreme of the number of necessary samples~\cite{M2k} for under-sampled signal recovery. Our algorithms thus start the first iteration by activating $M^{(0)}=2k$ sensors at locations randomly chosen from the set $\mathbb{L}$ to take samples. Our simulation study on the impact of the starting number of sensors at the first iteration also reveals that $2k$ is the optimal choice in terms of the final total sensing resources needed.


In our proposed adaptive schemes, instead of distributing the sensing resources evenly to randomly sample the signal vector and repeating the same sensing process over time, in each following iteration, we propose to just take a few new samples around the vector entries detected to have a higher chance of being nonzero from the reconstruction result of the previous iteration. Then the new samples combined with existing samples, as in Equation~(\ref{eq:MatrixReConstruction}), are applied to reconstruct the signal vector.

We use $\hat{s}^{(i)}$ to denote the reconstructed signal after the $i^{th}$ round of sensing. As we start from sensing with the number of sensors smaller than necessary for CS recovery, there exists inaccuracy in getting each intermediate $\hat{s}^{(i)}$. Although neither the positions nor the values of the nonzero entries of $\hat{s}^{(i)}$ may be accurate, our preliminary studies
 indicate that the actual nonzero entries of the original vector are close to the nonzero positions indicated by $\hat{s}^{(i)}$. In our signal detection example, the signal sources are possibly located in the region close to the grids corresponding to the nonzero positions of $\hat{s}^{(i)}$. So by "moving" sensing focus (i.e., selecting sensors at the desired locations) towards the estimated locations of the signal sources step by step, the algorithms will improve the reconstruction results until the positions and the values of the nonzero entries no longer change. This way we can find the accurate positions and energy levels of the signal sources. This also helps to increase the sensing efficiency and reduce the number of sensors needed. This is the fundamental principle of our adaptive algorithms.

The learning process iterates until the recovered result has reached a desired quality or cannot be further improved. Our results in Section~\ref{Section:Simulation} will show that the total number of sensors used in the overall adaptive process is much less than that needed for a single-time conventional CS recovery for achieving the same level of accuracy.

One thing to note is that $\hat{s}^{(i)}$ may contain many nonzero entries with very small values, which are insignificant and tedious to consider as the possible positions of targets. It has to be modified before each next iteration. To be specific, first all the negative entries of $\hat{s}^{(i-1)}$ will be set to zero. Then for the positive entries, all those with values below $\alpha\%$ that of the largest positive entry value in $\hat{s}^{(i-1)}$ will be set to zeros. The impact of $\alpha$ on performance is inspected in the simulations.

Based on the same adaptive principle, we propose two different algorithms on choosing new sampling locations in the next round for the signal strength detection problem, depending on signal source location distributions.

\subsection{Individual Chasing}
\label{Individual}
If there is no knowledge on the locations or distribution features of signal sources, it is reasonable to assume they are randomly distributed in the monitoring area. In this case, we will adapt the positions for sampling towards the estimated location of each individual signal source given by the last iteration.

In the $i^{th}$ iteration, according to the previous reconstruction vector $\hat{s}^{(i-1)}$, for each of its nonzero entry $\hat{s}^{(i-1)}(n)$, a sensor in set $\mathbb{L}$ whose location is closest to grid $n$ is selected to take a sample if a measurement is not already taken there before. It is best to choose the sensor right inside grid $n$ for sampling. However if there is no sensor located in the grid $n$, another sensor in the sensor location set $\mathbb{L}$ with the smallest euclidian distance to grid $n$ will be selected.
After each non-zero position $n$ is ensured to have one sampling in the corresponding grid, the $l_1$-minimization process is invoked to get the reconstruction $\hat{s}^{(i)}$ based on the combined samples $y$ and combined $\Phi$ in Equation~(\ref{eq:MatrixReConstruction}). The reconstruction result $\hat{s}^{(i)}$ is fed into Algorithm~\ref{Alg:TerminationCheck} for termination condition check to determine whether the algorithm should end or continue with more iterations. Algorithm~\ref{Alg:IndividualChasing} outlines the details of Individual Chasing in each iteration.

\begin{figure}[h]
\centering
\includegraphics[width=3in]{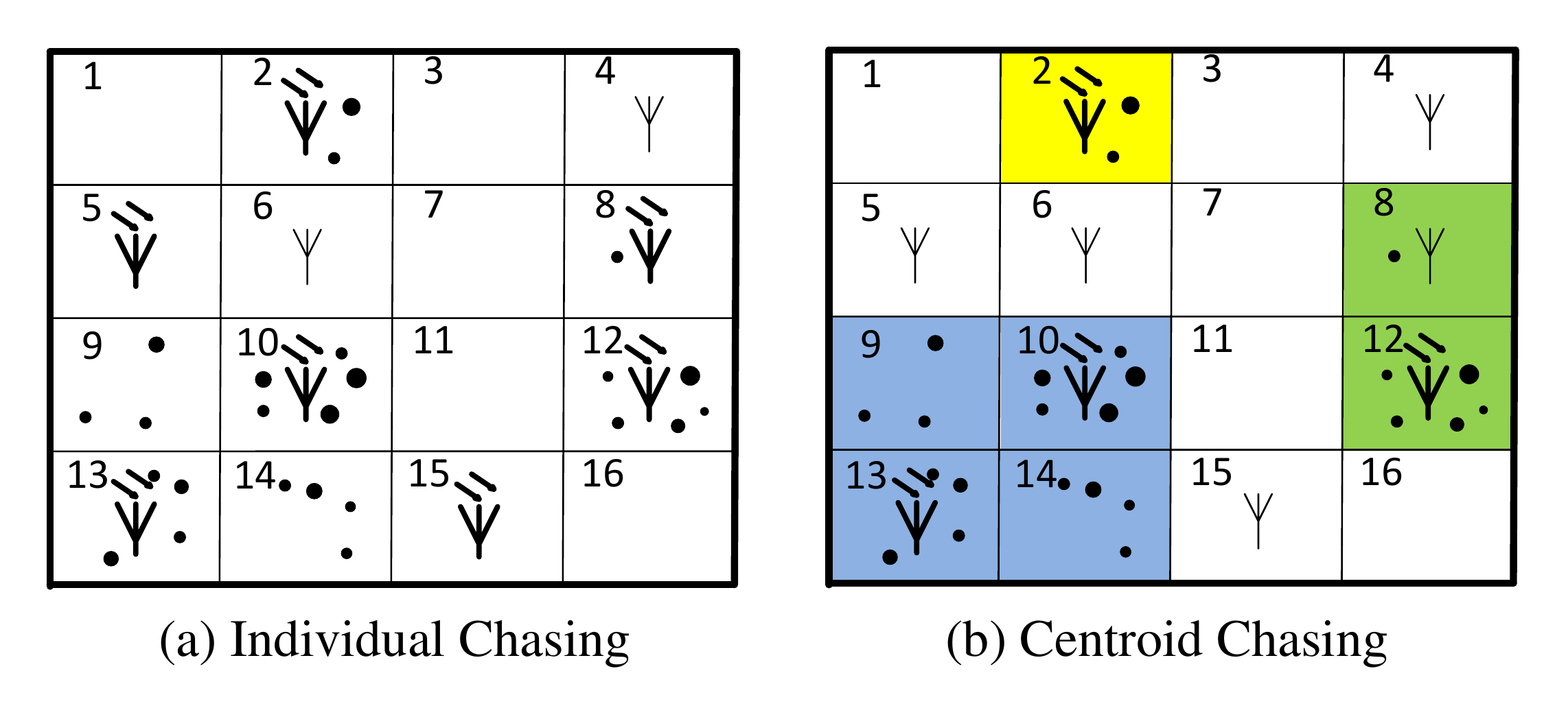}
\caption{Illustration of Individual Chasing and Centroid Chasing.}
\label{fig:Algorithms}
\end{figure}

Figure~\ref{fig:Algorithms}-(a) shows the sensor locations when the Individual Chasing algorithm terminates. It can be observed that sampling measurements have been taken at each grid with signal source that also has a sensor inside. For the grids with signal sources but without sensors deployed, samples are taken at the closest grids, i.e. samples have been taken by sensors in grid 5 and 15 for nearby signals in grid 9 and 14.

The Individual Chasing scheme adapts well when signal sources are uniformly distributed in the monitoring field. More importantly, the simulations in Section \ref{Section:Simulation} demonstrate that the Individual Chasing algorithm converges fast with superior recovery accuracy regardless of how the signal source locations are distributed.

\begin{algorithm}
\caption{Individual Chasing}
\label{Alg:IndividualChasing}
\begin{algorithmic}[1]
\STATE In the $i^{th}$ iteration:
\FOR {each nonzero position $n$ of $\hat{s}^{(i-1)}$}
\STATE find a grid position $p$ in $\mathbb{L}$ with the smallest euclidian distance to grid $n$.
\IF{no measurement has previously been taken at grid $p$}
\STATE take sampling at grid $p$'s sensor.
\ENDIF
\ENDFOR
\STATE combine new samples with existing ones for $y$.
\STATE do $l_1$-minimization on $y$ and $\Phi$ to get $\hat{s}^{(i)}$.
\STATE call Algorithm~\ref{Alg:TerminationCheck} to check the termination condition.
\IF{algorithm does not terminate in this iteration}
\STATE $i = i+1$, go back to Line 1 and start the next iteration.
\ENDIF
\end{algorithmic}
\end{algorithm}

\begin{algorithm}
\caption{Termination Condition Check}
\label{Alg:TerminationCheck}
\begin{algorithmic}[1]
\IF{the nonzero positions of $\hat{s}^{(i)}$ are all the same as $\hat{s}^{(i-1)}$}
\IF{the numeric difference of each nonzero value between $\hat{s}^{(i)}$ and $\hat{s}^{(i-1)}$ is smaller than a percentage threshold $\Delta^*$ of $\hat{s}^{(i-1)}$}
\STATE reconstruction process terminates in this iteration.
\ELSE
\STATE continue with the next iteration of reconstruction.
\ENDIF
\ELSE
\STATE continue with the next iteration of reconstruction.
\ENDIF
\STATE \small{$*$ \textit{the choice of $\Delta$ is given in Section\ref{Section:Simulation}}}
\end{algorithmic}
\end{algorithm}

\subsection{Centroid Chasing}
\label{Centroid}
In some scenarios, the signal sources may locate closely and form clusters. The clustering patterns of signal sources may be exploited to guide the sampling locations to facilitate the finding of signal sources with even fewer sensor measurements. For example, in an urban area, a region with high noise often indicates the existence of some events, such as street concert, parade, or riot. In this case, many sound sources may stay closely, and there is a need to detect the high noise areas to identify the events and protect against potential harm to the community.
The microphones of cell phones can be exploited to collaboratively detect the strength and location of these sound sources.

The Centroid Chasing scheme initializes similarly as Individual Chasing by randomly activating $M^{(0)}=2k$ sensors from the set $\mathbb{L}$ to start the first iteration. Algorithm~\ref{Alg:CentroidChasing} details the actions for each consecutive iteration. Let $\mathbb{T}^{(i-1)}$ denotes the set of grid numbers that correspond to the non-zero positions of the reconstructed $\hat{s}^{(i-1)}$, which indicate the possible grid positions where signal sources may reside estimated in the $i^{th}$ round. In the $i^{th}$ iteration, grid numbers in $\mathbb{T}^{(i-1)}$ are grouped into $C$ clusters based on their mutual euclidian distances, with $\mathbb{T}^{(i-1)}_t$ denoting the set of grids belonging to the $t^{th}$ ($t=1, ..., C$) cluster. The optimal clustering criteria is given later. The minimum rectangular region that covers all the grids of the $t^{th}$ cluster is called the cluster region of the $t^{th}$ cluster, and its size $R^{(i-1)}_t$ is the number of grids it covers. For each cluster, based on its signal target density defined as $|\mathbb{T}^{(i-1)}_t| / R^{(i-1)}_t$, $M^{(i-1)}_t$ closest sensors (these sensors could be either within or outside of the rectangular cluster region) will be selected for samplings at this iteration, where

\begin{equation} M^{(i-1)}_t = |\mathbb{T}^{(i-1)}_t| \cdot (1- \frac{|\mathbb{T}^{(i-1)}_t|}{R^{(i-1)}_t} ).
\label{eq:CC}\end{equation}
Equation~(\ref{eq:CC}) reflects the tendency to put less sensing resource in a cluster region when the target sources are denser. The number of sampling positions to be considered for adding at each iteration is controlled to be less than that of the Individual Chasing method, which is $|\mathbb{T}^{(i-1)}_t|$, by multiplying it with the density which is a value smaller than 1. This feature, however, may result in more iterations needed thus more sensors involved than the Individual Chasing algorithm when the signals are sparsely distributed and poorly clustered. The performances of two algorithms are compared under different scenarios in the simulation. Again, the new measurements obtained at the current iteration will be combined with existing samples for CS reconstruction, which will be checked by termination condition afterwards.

\begin{algorithm}
\caption{Centroid Chasing}
\label{Alg:CentroidChasing}
\begin{algorithmic}[1]
\STATE In the $i^{th}$ iteration:
\STATE group the grids in $\mathbb{T}^{(i-1)}$ into $C$ clusters.
\FOR {each of the cluster $\mathbb{T}^{(i-1)}_t$}
\STATE find the $M^{(i-1)}_t$ sensors in $\mathbb{L}$ that are closest in euclidian distance to the center of the $t^{th}$ cluster region.
\IF{any of these $M^{(i-1)}_t$ chosen sensor grids has not sampled before}
\STATE activate the sensors at these grids.
\ENDIF
\ENDFOR
\STATE combine new samples with existing ones for $y$.
\STATE do $l_1$-minimization on $y$ and $\Phi$ to get $\hat{s}^{(i)}$.
\STATE call Algorithm~\ref{Alg:TerminationCheck} to check the termination condition.
\IF{algorithm does not terminate in this iteration}
\STATE $i = i+1$, go back to Line 1 and start the next iteration.
\ENDIF
\end{algorithmic}
\end{algorithm}

Specifically, Line 2 of Algorithm~\ref{Alg:CentroidChasing} clusters the possible signal sources whose positions are estimated by the nonzero positions of $\hat{s}^{(i-1)}$. This procedure starts at an initial empirical number of clusters $C=\sqrt{|\mathbb{T}^{(i-1)}|/2}$, where $|\mathbb{T}^{(i-1)}|$ is the total number of elements in the set to be clustered, as given in \cite{Kmean}. The optimal number of clusters to form is found by varying the number of clusters $C$ and picking the one under which the average distance between each grid position to the centroid of cluster it belongs to is minimal. After clustering, a smaller number of sensors can be tasked in the signal concentrated area to achieve even higher sensing efficiency.

Figure~\ref{fig:Algorithms}-(b) shows the sensing condition at the end of the Centroid Chasing algorithm. Grids with signals are grouped into 3 clusters. Within each cluster, a portion of the closest sensors are activated for sampling. Compared with Individual Chasing scheme on the left of the figure, fewer number of sensors are used given the signal sources have a nice clustering feature.

The Centroid Chasing algorithm is specially suitable for applications with knowledge of clustered distribution, which, nonetheless, does not prevent its applicability in general cases. The simulations in section \ref{SubSection:SimulationNumberOfSamples} show that Centroid Chasing algorithm outperforms the Individual Chasing algorithm when the signal sources are denser in clusters, and works almost equally efficient in general scenarios where the signal sources are sparsely distributed.

\subsection{\textbf{Local Optimum Avoidance - Random Exploration}}
\label{Local_Optimum}
Very unlikely but possible, adaptive algorithms could converge at the local optimum. In our problem, local optimum does not give accurate reconstruction at the algorithm termination time. In an iterative scenario, one can only look at the intermediate results to decide whether the iteration should stop. The Individual Chasing and Centroid Chasing algorithms will stop when consecutive iterations present no more change in the results. In order to avoid possible local optimum, we introduce an extra step called Random Exploration. To be specific, for both algorithms, when the Termination Condition Alg.\ref{Alg:TerminationCheck} is satisfied at certain iteration, we do not terminate the program immediately. Instead, we randomly pick up some sensors that have not been activated before to take samples, then re-enter the adaptive process and let it converge again. This random walk procedure is proved to work extremely well in our simulations.

%% file: Sec6_Simulation.tex
\section{Simulation and Performance Evaluation}
\label{Section:Simulation}
We evaluate the performance of our two proposed learning-based compressive sensing schemes through extensive simulations. Before showing our simulation results, we first introduce our performance metrics and simulation set-up.

\subsection{\textbf{Performance Metrics}}
\begin{itemize}
  \item \textbf{Reconstruction Error}: defined as the Sum of Absolute Difference (SAD) between the recovered and the original signal vector:
      \begin{equation}SAD=\|\hat{s} - s\|_{l_1} = \displaystyle\sum_{i=1}^{N} |\hat{s}_i - s_i|\end{equation}
       This measurement metric evaluates the accuracy of vector reconstruction. It reflects not only the degree of error due to the position mismatch of nonzero entries, but also the difference in magnitude for each unmatched entry.
  \item \textbf{Number of Sensors Needed ($M$)}: in our adaptive algorithms, new sensors may be added for sampling in each new iteration. The number of sensors needed for our algorithms are defined as the total number of sensors used during the whole sensing process, which is the least number of sensors needed in our performance studies to give $100\%$ reconstruction accuracy.
\end{itemize}

\subsection{\textbf{Simulation Set-Up}}

We simulate the general problem on forming the energy measurement map and compare the performances of our proposed schemes with peer works.
$dBm$ is adopted as the measurement unit of signal strength in our simulation. One grid could have multiple signal sources, and the overall signal strength for a grid is the aggregate of these signals. Being aware that the numeric scale of the nonzero entries of signal vector is not critical to the problem of compressive sensing recovery but actually the sparsity is, we assume a range of 30-500 $dBm$ for the possible aggregated signal strength inside any single grid. For every simulation run, signal sources are generated at random locations across the sensing region with the aggregate signal strength of a grid to be a random value selected from the 30-500 range.


The simulation is carried out with MATLAB. Our schemes improve the sensing performance benefitting from the adaptive chasing process, but do not rely on any adopted specific channel model as previously pointed out. In the simulation, we would like to use a specific radio channel model to test our schemes. We define the channel as the following. The strength at location $j$ for a signal source at location $i$ is roughly approximated as:
\begin{equation}\mathbb{P}_{ij}= \frac{P_i \mathbb{G}_{ij}}{d^\beta_{ij}}, \; where \; \mathbb{G}_{ij}= \mathbb{X}_{ij} + \mathbb{Y}_{ij} \cdot \mathrm{i} \label{eq:FadingModel}\end{equation}
$P_i$ is the signal strength at its source location $i$, which is essentially the aggregated strength of signal sources at grid $i$, i.e. $s_{i}$
The denominator represents the path loss due to the distance $d_{ij}$  between locations $i$ and $j$.
$\beta$ is the decaying factor with possible value in $[2.0, 5.0]$, depending on the environment. `$\mathrm{i}$' is the imaginary sign. $\mathbb{G}_{ij}$ is a complex random Gaussian variable with real and imaginary components both being independent and identically distributed zero-mean Gaussian variables $\mathbb{X}_{ij}, \mathbb{Y}_{ij} \sim \mathcal{N}(0,\sigma^2_0)$, which captures the Raleigh distribution for multi-path fast fading of the signal \cite{FadingModel1,FadingModel2}.
The variance $\sigma_0$ is set to 0.5 as in~\cite{TargetCounting} for fair comparison that follows. $\frac{ \mathbb{G}_{ij}}{d^\beta_{ij}}$ corresponds to $C_{ij}$ in the channel model of Equation~(\ref{eq:Psi}).

In the short duration of performing the sensing task, sensors (essentially cell phones in our application) will not have significant moving distance or move across grids.
For each sensing task, 400 sensors are randomly deployed in an area of $N=30\times30=900$ grids with at most one sensor inside one grid, as we can always select one sensor to participate in the sensing task when a grid has multiple sensors inside as discussed previously. The size of each square grid is set to 30 meter which can reasonably reflect the distance effect in signal propagation. Although the grid size determines the signal monitoring resolution, it does not have significant impact on the performance of our algorithms once the resolution requirement is given. There are $k (k << N)$ grids with signal sources at a given time instant. $k$ is the sparsity value, which is varied in different simulation studies.
The termination condition threshold $\Delta$ in Alg.\ref{Alg:TerminationCheck} is set to $5\%$ which will generally guarantee the recovery result to be accurate at the algorithm termination time with the recovery error in the order of $10^{-4}$ even for real valued signal vectors based on our preliminary studies.
To compare our two schemes, the positions of $k$ grids with signal sources can be either distributed in a clustered fashion, or uniformly distributed. To examine the reliability performance of our schemes, Gaussian White noise $N(0,\sigma^2)$ is added to the observed sample vector $y$ in some of the simulation runs, and SNR measure is exploited to quantify the noise strength. Each presented result is the average of many runs.

Our proposed two algorithms Individual Chasing and Centroid Chasing, also referred to as IC and CC from now on, do not depend on any specific CS reconstruction technique.
Thus we chose two fundamental and most prevalent types of work for performance evaluations-$l_1$ minimization based CS and greedy based CS. GMP~\cite{TargetCounting} provides a greedy based reconstruction algorithm for CS, and also exploits the received signal strength at different grid positions to help solve the target localization and counting problem. $l_1$-magic is a concise and dominant realization of $l_1$ minimization based CS scheme, which can be directly applied to and is thus worth comparing with our simulation scenario of signal strength vector reconstruction.


\subsection{\textbf{Parameter Study}}
\label{SubSection:SimulationParameterStudy}
\begin{figure}[htbp]
\centering
\includegraphics[width=3.6in]{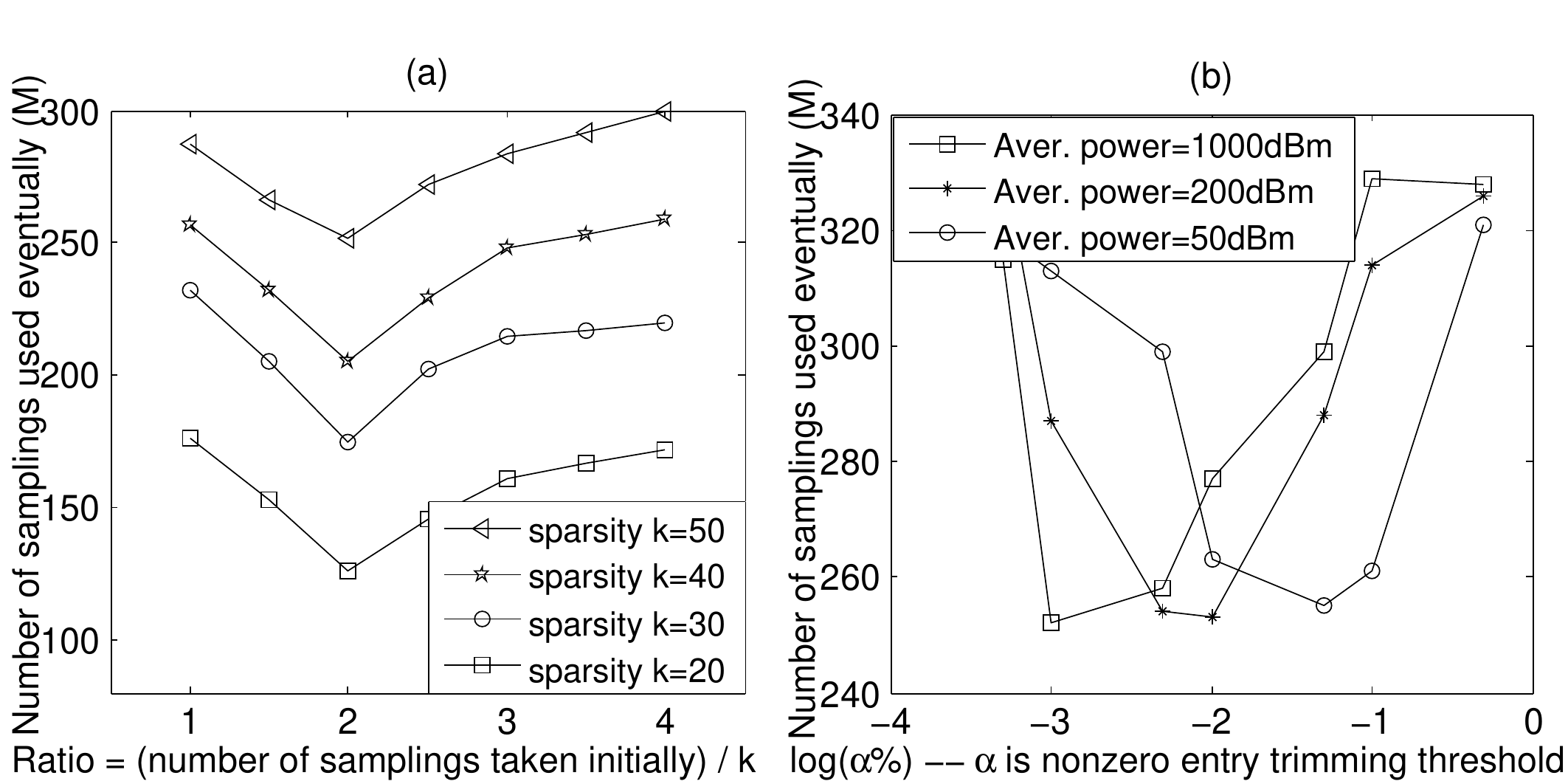}
\caption{(a) The number of sensors ultimately used vs. the number of sensors started with. (b) The number of sensors ultimately used vs. the trimming threshold $\alpha$}
\label{fig:ParameterStudy}
\end{figure}

For our adaptive algorithms, the total number of sensors used increases after each iteration. Naturally, the number of sensors taken in the first iteration would have impact on the final total number of sensors used. In Figure\ref{fig:ParameterStudy}-(a), we study the optimal starting number of sensors for our Individual Chasing algorithm. For each sparsity $k$, the optimal choice on the number of sensors to start with is clearly around $2k$. Intuitively, if we start with too few sensors, the intermediate recovered vectors will be very inaccurate and result in much more iterations thus more sensors involved in total; however starting with too many sensors at the beginning would provide more sensing resources than needed which again causes more sensors in total. Therefore the number of sensors to start with for both Individual Chasing and Centroid Chasing algorithms will be set to twice the value of $k$ in the following simulations if not otherwise specified.

In Section \ref{Section:MainScheme}, we mentioned the trimming threshold $\alpha$, based on which a portion of insignificant nonzero entries of the recovered vector are set to `0' before the next iteration in order to shrink the target range of nonzero entries to work on. In Figure~\ref{fig:ParameterStudy}-(b), we fix $k$ at 50, but vary the average signal source transmitting power (i.e. magnitude of each vector entry) at three distinct levels-1000, 200 and 50 dBm, and investigate the number of sensors needed for accurate recovery using each different trimming threshold. Since the range of $\alpha$ picked spans over several orders of magnitude, we apply base-10 logarithm on $\alpha\%$ as the x-axis. It can be clearly observed that for each setting of signal vector magnitude, there is an optimal range of $\alpha$ that gives the minimum number of sensors needed. Out of this range, the performance deteriorates and quickly reach a steady level.
This is because only when the $\alpha\%$ is appropriately chosen that not too many components of the vector are wiped out nor too many insignificant ones are preserved will the reconstruction process requires the least number of sensors.

With the aggregate signal strength ranging between 30-500 dBm, the average signal strength in our simulation is close to 200 dBm. According to the figure, $1\%$ is adopted as the default trimming threshold in the simulations that follow.

\subsection{\textbf{Number of Sensors Needed}}
\label{SubSection:SimulationNumberOfSamples}
\begin{figure}[htbp]
\centering
\includegraphics[width=3.6in]{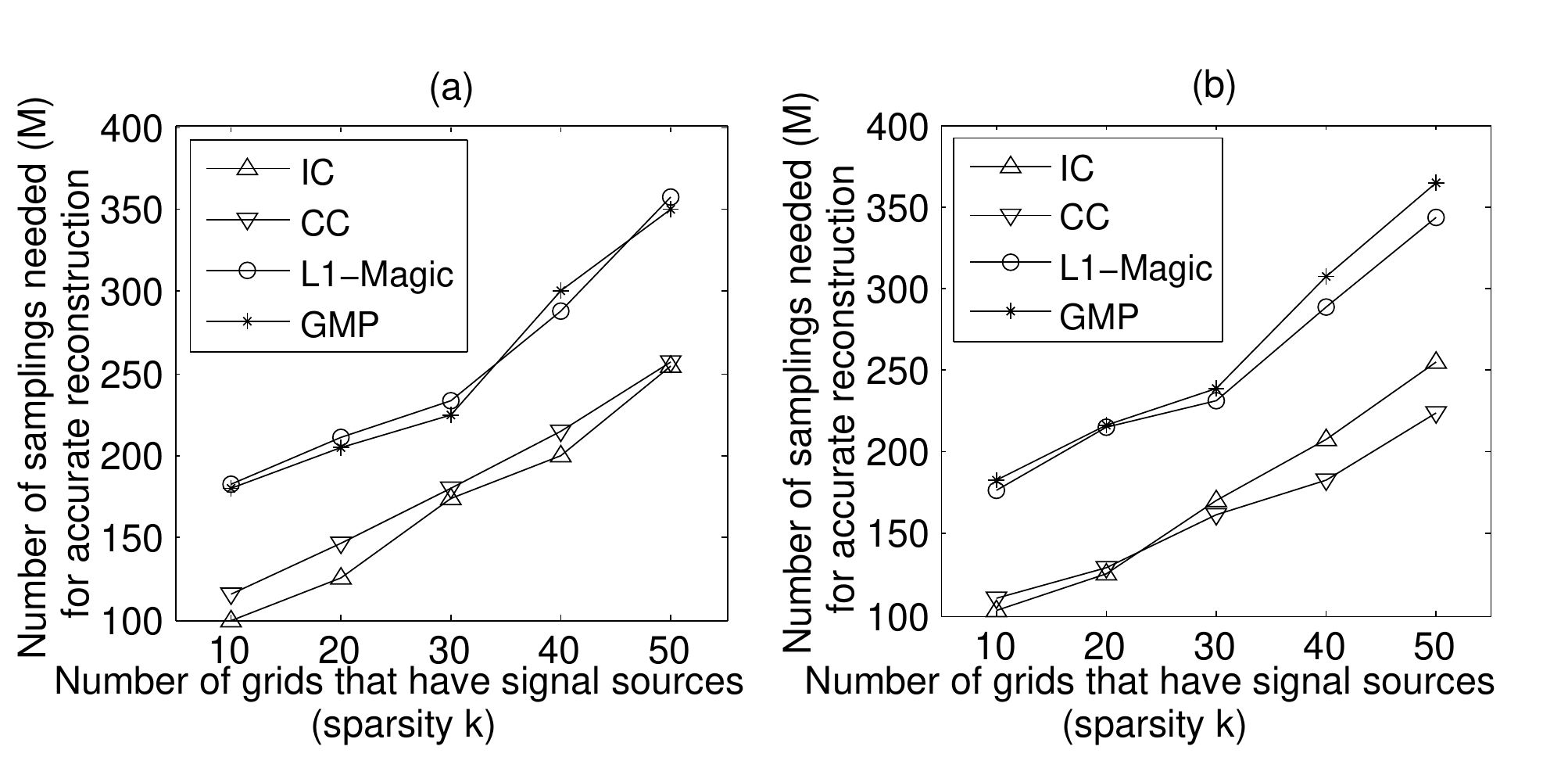}
\caption{(a) scattered signal sources. (b) clustered signal sources.}
\label{fig:NumberOfSensorsNeeded}
\end{figure}

Pursuing a minimum number of sensors needed for an accurate signal reconstruction is the major challenge and research focus in the closely related research fields. We evaluate the minimum number of sensors needed for accurate signal vector reconstruction (zero reconstruction error)
under different levels of signal sparsity for each scheme in Figure \ref{fig:NumberOfSensorsNeeded}. As expected, the number of sensors needed increases as $k$ grows for all the algorithms. Particularly, Figure~\ref{fig:NumberOfSensorsNeeded}-(a) is under the scenario where the signal sources are randomly uniformly distributed across the network grids. IC performs slightly better than CC as expected. The clustering function of CC is not effective when the signal sources are uniformly distributed, which leads to more iterations to converge and more sensors needed. Compared to GMP, IC requires $45\%$ fewer sensors when $k$ is small, and about $23\%$ fewer sensors when $k$ gets bigger. IC requires $46\%-25\%$ fewer sensors than $l_1$-magic for different $k$.

In Figure~\ref{fig:NumberOfSensorsNeeded}-(b), the signal sources are distributed in clustered fashion across grids, which benefits the clustering process of CC algorithm.
 Thus CC is observed to require fewer sensors than IC as the number of signal sources exceeds certain value. While the performance difference between IC and CC is small, CC requires $40\%-33\%$ fewer sensors than GMP, and $40\%-30\%$ fewer than $l_1$-magic, which are big improvements. In general, both IC and CC work extremely well no matter the signal sources are concentrated or scattered, and far outperform the other two schemes under the same $k$.

Given the performance difference between IC and CC is small and both algorithms follow the same principle, we only study and compare IC with the other two schemes and assume the signal sources are randomly and uniformly distributed in the simulations that follow.

\subsection{\textbf{Reconstruction Error}}
\subsubsection{\textbf{Convergency Study}}
\label{SubSection:SimulationConvergency}
\begin{figure}[htb]
\centering
\includegraphics[width=3.6in]{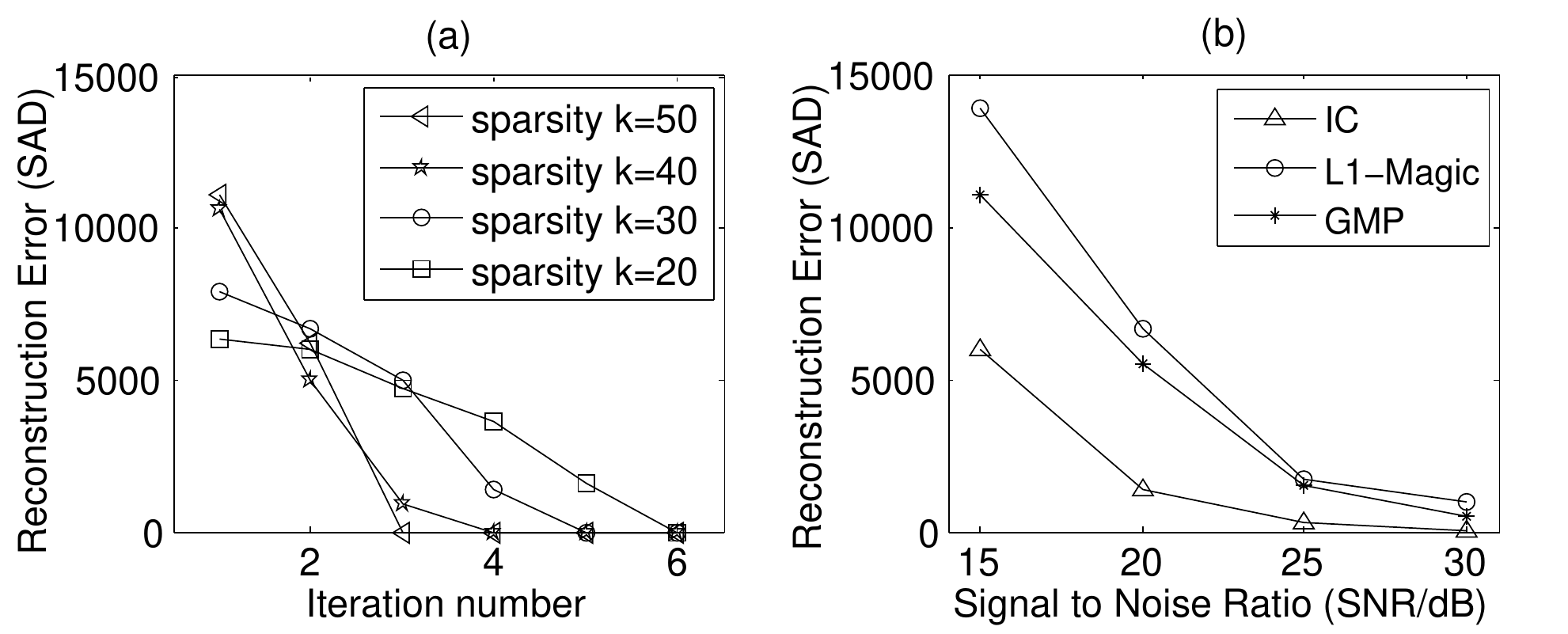}
\caption{(a) Convergency study of Individual Chasing algorithm. (b) The reconstruction error comparison due to noise under the same $k=50$ and $M=250$.}
\label{fig:ReconstructionError}
\end{figure}

We study the convergency of IC in Figure~\ref{fig:ReconstructionError}-(a). It is clear that under all $k$, IC is able to converge within 3-6 iterations to get accurate reconstruction with 0 error, and it exhibits a rather steady (approximate-linear) improvement in reducing the reconstruction error in each iteration. It converges faster for larger $k$. This is due to the fact that we initialize $2k$ number of sensors for the optimal performance. With a larger $k$, there are more samples taken at the beginning, therefore it needs fewer iterations to get enough overall sensor samples for accurate recovery.

\subsubsection{\textbf{Performance Under Noise}}
\label{SubSection:SimulationNoise}
The adaptive algorithms can always find the accurate reconstruction given enough iterations and proper handling with local optimum avoidance. However in a noisy environment where the sample $y$ is contaminated, the final reconstructed result could be different from the actual signal vector.

The reconstruction errors due to sampling noise for different algorithms are compared in Figure~\ref{fig:ReconstructionError}-(b). The reconstruction error reduces as signal-to-noise ratio increases for all three schemes. Under the same sensing condition of $k=50$ and $M=250$, at each SNR level, IC gives much more accurate result than the other two. In the worst scenario with the strongest noise at 15dB SNR in our test setting, the reconstruction error for IC is approximately $(14000-6000)/14000=57\%$ smaller than that of $l_1$-magic. GMP is slightly more accurate than $l_1$-magic under the same sensing setting, because it enumerates all possible values for each possible nonzero position of the vector at the cost of higher computational overhead.

%% file: Sec7_Conclusion.tex
\section{Conclusion}
\label{Section:Conclusion}
Rather than using conventional random sampling in compressive sensing, we observe and theoretically prove that by learning from the sensing results and adjusting the structure of the sensing matrix adaptively, the number of samples needed for high-quality signal recovery in compressive sensing can be significantly lower than that required by the basic $l_1$ minimization solution. We propose two learning-based adaptive algorithms, Individual Chasing and Centroid Chasing, for different signal source distribution scenarios. Both schemes adaptively concentrate sensing resources to proper signal subspace towards better acquisition of signals, and do not depend on any specific CS reconstruction methods. The instantiation of our algorithms to solve the general signal strength sensing problem with distance fading can be conveniently generalized to various similar applications.
Extensive simulations demonstrate that our algorithms can achieve as much as $57\%$ more accurate signal recovery under noisy conditions, and require up to $46\%$ fewer sensors than state-of-the-art related works.
